\documentclass[aps,twocolumn,prl,superscriptaddress,amsmath,tightenlines,nofootinbib]{revtex4}
\usepackage{epsfig,graphicx,times}
\usepackage{amstext}
\usepackage{amsmath}            
\usepackage{amssymb}            
\usepackage{graphicx}           
\usepackage{latexsym}
\usepackage{bm}
\usepackage{color}

\def \I{{\rm i}}

\begin{document}

\title{Controllable microwave three-wave mixing via
a single three-level superconducting quantum circuit}

\author{Yu-xi Liu}
\affiliation{Institute of Microelectronics, Tsinghua
University, Beijing 100084, China} \affiliation{Tsinghua
National Laboratory for Information Science and Technology
(TNList), Tsinghua University, Beijing 100084, China}
\affiliation{CEMS, RIKEN, Saitama 351-0198, Japan}

\author{Hui-Chen Sun}
\affiliation{Institute of Microelectronics, Tsinghua
University, Beijing 100084, China} \affiliation{CEMS, RIKEN,
Saitama 351-0198, Japan}

\author{Z. H. Peng}
\affiliation{CEMS, RIKEN, Saitama 351-0198, Japan}

\author{Adam Miranowicz}
\affiliation{Faculty of Physics, Adam Mickiewicz University,
61-614 Pozna\'n, Poland} \affiliation{CEMS, RIKEN, Saitama
351-0198, Japan}

\author{J. S. Tsai}
\affiliation{CEMS, RIKEN, Saitama 351-0198, Japan}
\affiliation{NEC Green Innovation Research Laboratories,
Tsukuba, Ibaraki 305-8501, Japan}

\author{Franco Nori}
\affiliation{CEMS, RIKEN, Saitama 351-0198, Japan}
\affiliation{Physics Department, The University of Michigan,
Ann Arbor, Michigan 48109-1040, USA}
\date{\today}

\begin{abstract}
Three-wave mixing in second-order nonlinear optical processes
cannot occur in atomic systems due to the electric-dipole
selection rules. In contrast, we demonstrate that second-order
nonlinear processes can occur in a superconducting quantum circuit
(i.e., a superconducting artificial atom) when the inversion
symmetry of the potential energy is broken by simply changing the
applied magnetic flux. In particular, we show that difference- and
sum-frequencies (and second harmonics) can be generated in the
microwave regime in a controllable manner by using a single
three-level superconducting flux quantum circuit (SFQC). For our
proposed parameters, the frequency tunability of this circuit can
be achieved in the range of about $17$ GHz for the sum-frequency
generation, and around $42$ GHz (or $26$ GHz) for the
difference-frequency generation. Our proposal provides a simple
method to generate second-order nonlinear processes within current
experimental parameters of SFQCs.
\end{abstract}

\maketitle \pagenumbering{arabic}

Nonlinear optical effects have many fundamental applications
in quantum electronics, atom optics, spectroscopy, signal
processing, communication, chemistry, medicine, and even
criminology. These phenomena include optical Raman
scattering, frequency conversion, parametric amplification,
the Pockels and Kerr effects (i.e., linear and nonlinear
electro-optical effects), optical bistability, phase
conjugation, and optical solitons~\cite{Shen-book,Boyd-book}.
Three-wave mixing (including the generations of the
sum-frequency, difference-frequency, and second harmonics)
and four-wave mixing are important methods to study nonlinear
optics. It is well-known that materials without inversion
symmetry can exhibit both second- and third-order
nonlinearities. However, materials with inversion symmetry
usually exhibit only third-order nonlinearities. Thus,
three-wave mixing (which requires the second-order
nonlinearity) cannot occur in atomic systems with
well-defined inversion symmetry, because the electric-dipole
transition selection rules produce a zero
signal~\cite{Shen-book} with mixed frequencies. Although
chiral molecular three-level systems without inversion
symmetry can be used to generate three-wave mixing in the
microwave
domain~\cite{Three-wave-1977a,Three-wave-1977b,Molecular-book,Patterson-PRL},
such wave mixing cannot be tuned because the energy structure
of the systems is fixed by nature.

Recently, superconducting charge, flux, and phase quantum circuits
based on Josephson junctions have been extensively explored as
basic building blocks for solid-state quantum information
processing~\cite{Clarke-review,Girvin-review,You-Review1,You-Review2}.
These circuits can also be considered as artificial
atoms~\cite{You-Review1,Buluta11}. In contrast to natural atoms,
the quantum energy structure and the potential energy of these
artificial atoms can usually be tuned by external parameters.
Thus, they can possess new features and can be used to demonstrate
fundamentally new phenomena which cannot be found in natural
three-level atoms. For example, with the tunable potential energy
of superconducting flux quantum circuits (SFQCs) by varying the
bias magnetic flux, three-level (qutrit) SFQCs can have a
$\Delta$-type (cyclic) transition~\cite{liu2005}. Two-level SFQCs
are also known as superconducting flux qubits~\cite{Mooij1999}.
Three-level SFQCs (i.e., superconducting flux qutrits) can be used
to demonstrate the coexistence of single- and
two-photons~\cite{liu2005,Gross}, which does not occur in natural
three-level atomic systems with electric-dipole interaction. Such
$\Delta$-type atoms can also be used to cool quantum
systems~\cite{You-cool}, or generate microwave
single-photons~\cite{You-single}.

In solid-state quantum information processing, microwave signals
are usually employed for measuring and controlling the qubits.
Moreover, these signals can also be used to detect the motion of
nanomechanical resonators~\cite{Nanomechanical} and to read out
the spin information in nitrogen-vacancy centers in
diamonds~\cite{NV-R1}. Therefore, the controllable generation,
conversion and amplification of microwave signals play a very
important role in solid-state quantum information processing. The
generation of microwave Fock's
states~\cite{Liu-singlephoton,Single-photon,UCSB1}, superpositions
of different Fock's states~\cite{UCSB2}, squeezed
states~\cite{Wallraff}, nonclassical microwave~\cite{delsing} and
giant Kerr nonlinearities~\cite{Giant1,Giant2} have been studied
in the microwave domain via circuit quantum electrodynamics
(QED)~\cite{Clarke-review,Girvin-review,You-Review1}. Microwave
parametric amplification~\cite{Schackert} has also been studied by
using three-wave mixing~\cite{Roch-PRL} in superconducting
circuits with four Josephson junctions. Different from
Ref.~\cite{Roch-PRL}, here we propose another method to generate
microwave three-wave mixing, including the generation of the sum-
and difference-frequencies in a controllable way via a tunable
single SFQC. This method also applies for
phase~\cite{Martinis,Hakonen1,Hakonen2} and
transmon~\cite{transmon} qutrits. In our proposal, such three-wave
mixing can be switched off at the optimal point by the bias
magnetic flux. We also discuss the possibility for the generations
of second harmonics and zero-frequency using SFQCs.

\begin{figure}
\epsfig{file=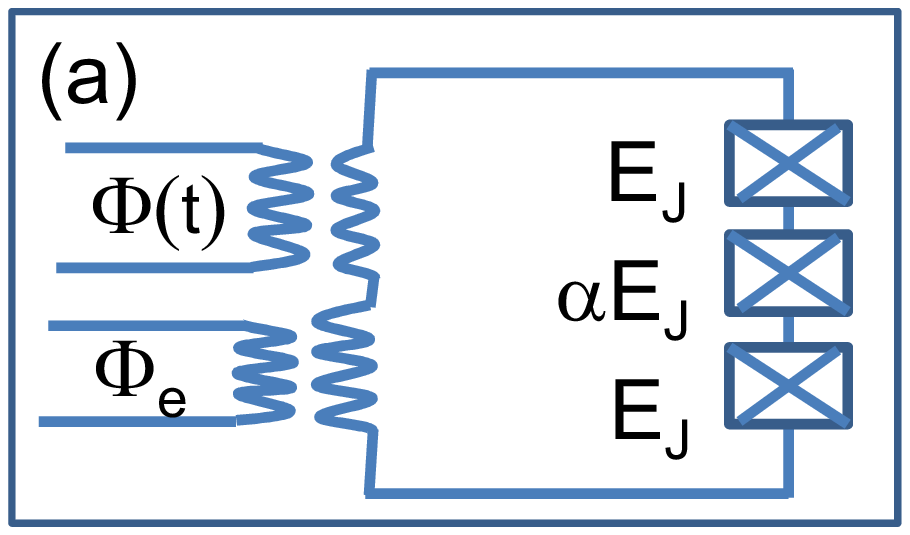,width=8.8cm}
\epsfig{file=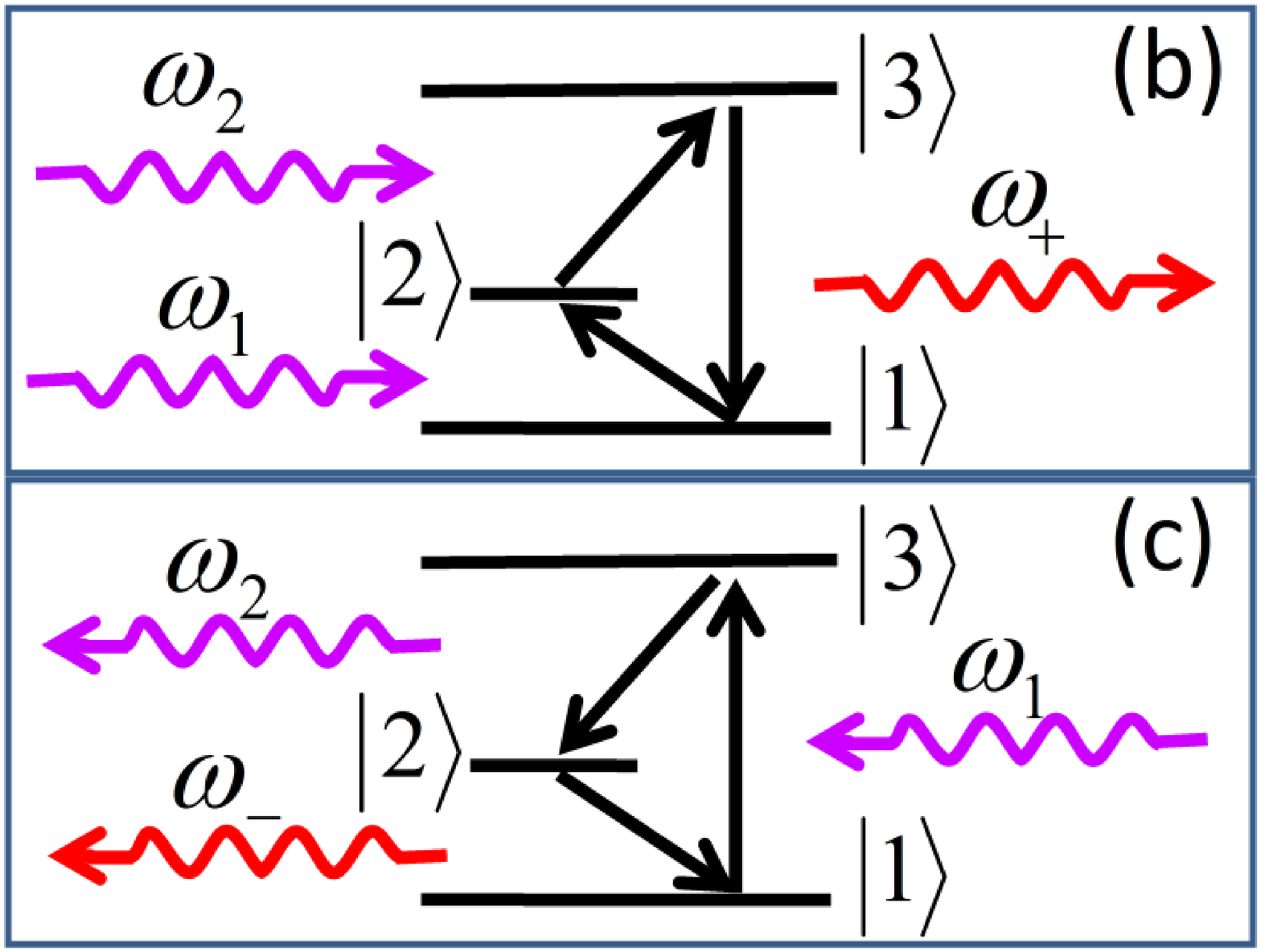,width=8.8cm}
\caption{(a) Schematic diagram for a SFQC with three
Josephson junctions biased by a magnetic flux $\Phi_{e}$ and
also driven by the magnetic flux
$\Phi(t)=\sum_{l}\Phi(\omega_{l})\exp(-\I\omega_{l}t)$ with
different frequencies $\omega_{l}$, which are specified  in
panels (b,c); $E_J$ is the Josephson energy, and $0.5<\alpha<
1$. (b) A three-level (qutrit) SFQC, which can be considered
as an artificial atom with $\Delta$-type (cyclic) transitions
driven by the external magnetic flux $\Phi(\omega_{1})$
[$\Phi(\omega_{2})$] with frequency $\omega_{1}$
($\omega_{2}$) to induce the transition between the energy
levels $|1\rangle$ and $|2\rangle$ ($|2\rangle$ and
$|3\rangle$), results in the generation of the output signal
with the sum-frequency $\omega_{+}$; (c) Same as in panel (b)
but for the flux $\Phi(\omega_{1})$ inducing the transition
between the energy levels $|1\rangle$ and $|3\rangle$, leads
to the generation of the output signal with the
difference-frequency $\omega_{-}$.} \label{fig1}
\end{figure}

\section*{Model}

To be specific, our study below will focus on three-level SFQCs,
also called a qutrit or three-level qudit. However, our results
can also be applied to phase and transmon qutrits. As shown in
Fig.~1(a), a SFQC consists of a superconducting loop interrupted
by three Josephson junctions and controlled by a bias magnetic
flux $\Phi_{e}$. The Josephson energies (capacitances) of the two
identical junctions and the smaller one are $E_{J}$ ($C_{J}$) and
$\alpha E_{J}$ ($\alpha C_{J}$) with $0.5<\alpha<1$, respectively.
If we assume that the SFQC is driven by the external
time-dependent magnetic flux
$\Phi(t)=\sum_{l}\Phi(\omega_{l})\exp(-\I\omega_{l}t)$ with
frequencies $\omega_{l}$, then we can describe the system by this
Hamiltonian
\begin{equation}\label{eq:1}
H=-\frac{\hbar^2}{2M_{p}}\frac{\partial^2}{\partial
\varphi_{p}^2}-\frac{\hbar^2}{2M_{m}} \frac{\partial^2}{\partial
\varphi_{m}^2}+U(\varphi _{p},\varphi _{m},f)+V(t)
\end{equation}
with $M_{p}=2C_{J}[\Phi_{0}/(2\pi)]^2$ and
$M_{m}=M_{p}(1+2\alpha)$. The potential energy is
\begin{eqnarray}
  U(\varphi _{p},\varphi _{m},f) &=&
2E_{J}(1-\cos \varphi _{p}\cos \varphi _{m})\nonumber \\
&&+\alpha E_{J}\left[1-\cos \left(2\pi
f+2\varphi_{m}\right)\right], \label{eq_U}
\end{eqnarray}
with phases $\varphi_{p}=(\phi_{1}+\phi_{2})/2$ and
\begin{eqnarray}
  \varphi_{m}=\frac{1}{2}(\phi_{2}-\phi_{1})+\frac{2\pi \alpha
}{2\alpha+1}\frac{\Phi(t)}{\Phi_{0}}, \label{eq_phi}
\end{eqnarray}
where $\phi_{1}$ and $\phi_{2}$ are the gauge-invariant phases of
the two identical junctions (see Fig.~1). Here
$f=\Phi_{e}/\Phi_{0}$ is the reduced magnetic flux, and
$\Phi_{0}=h/(2e)$ is the flux quantum. The interaction between the
SFQC and the time-dependent magnetic flux is described by $V(t)=I
(\varphi_{p},\varphi_{m},f)\Phi(t)$, with the supercurrent
\begin{equation}
I (\varphi_{p},\varphi_{m},f)=\frac{\alpha\,I_{0}}{2\alpha+1}
\left[\sin\left(2\pi f+2
\varphi_{m}\right)-2\sin\varphi_{m}\cos\varphi_{p} \right]
\label{eq_I}
\end{equation}
inside the superconducting loop~\cite{liu2014,liu2006} and
$I_{0}=2\pi E_{ J}/\Phi_{0}$. The supercurrent $I\equiv I
(\varphi_{p},\varphi_{m},f)$ and the external magnetic flux
$\Phi(t)$ are equivalent to the electric dipole moment operator
and time-dependent electric field of the electric dipole
interaction in  atomic systems. It is obvious that $U(\varphi
_{p},\varphi _{m},f)$ in Eq.~(\ref{eq:1}) can be tuned by the bias
magnetic flux $\Phi_{e}$. We have shown that one of two flux quits
cannot work at the optimal point when both qubits are directly
coupled through their mutual inductance~\cite{liu2006}, because of
its selection rules~\cite{liu2005,liu2014}. Such problem can be
solved by introducing a coupler (e.g., see,
Refs.~\cite{liu2007-PRB, Harrabi2009,Sahel2008}).

We have shown~\cite{liu2005} that three-level SFQCs have
$\Delta$-type (cyclic) transitions among the three lowest energy
levels $|i\rangle$ when the inversion symmetry of the potential
energy is broken, otherwise it has a cascade transition. Under the
three-level approximation of SFQCs, Eq.~(\ref{eq:1}) becomes
\begin{equation}\label{eq:5}
  H_{T}=\sum_{i=1}^{3}E_{i}|i\rangle\langle i|+V_{T}(t),
\end{equation}
where $E_{i}$ ($i=1,\,2,\,3$) are three eigenvalues corresponding
to the three lowest eigenstates $|i\rangle$ of Eq.~(\ref{eq:1})
with $V(t)=0$. With this three-level approximation of SFQCs, the
interaction Hamiltonian $V_{T}(t)$ in Eq.~(\ref{eq:5}) can be
generally written as
\begin{eqnarray}\label{eq_V_T}
V_{T}(t)=\left[\sum_{i, j=1, i<j}^{3}I_{ij}(f)\sigma_{ij}+{\rm
H.c.}\right]\Phi(t),
\end{eqnarray}
with operators $\sigma_{ij}=|i\rangle\langle j|$ and matrix
elements $I_{ij}(f)\equiv\langle
i|I(\varphi_{p},\varphi_{m},f)|j\rangle$  dipole-like moment
operator. Here, the longitudinal coupling
$\sum_{i=1}^{3}I_{ii}(f)\sigma_{ii}\Phi(t)$ between the
three-level SFQC and the time-dependent magnetic flux is neglected
even though the reduced magnetic flux is not at the optimal
point., i.e., $f\neq0.5$ . We note that $f=0.5$ is called as the
optimal point or the symmetry point~\cite{Mooij1999}, where the
influence of flux noise is minimal. When the relaxation and
dephasing of the three-level SFQC are included, the dynamics can
be described by the master equation
\begin{eqnarray}\label{eq:2}
\dot{\rho}(t)&=&\frac{1}{\I\hbar}[H_{T}, \rho]+\frac{1}{2}\sum_{i=2}^{3}\gamma_{ii}(2\sigma_{ii}\rho\sigma_{ii}-\sigma_{ii}\rho-\rho\sigma_{ii}-\overline{\rho}_{ii})\nonumber\\
&-&\frac{1}{2}\sum_{l=1}^{3}\sum_{i<j}\gamma_{ij}
\left[(\sigma_{jj}\rho-\overline{\rho}_{jl}\sigma_{jl})-(\rho\sigma_{jj}-\overline{\rho}_{lj}\sigma_{lj})\right]\nonumber\\
&+&\sum_{i<j}\gamma_{ij}
\sigma_{ij}(\rho-\overline{\rho}_{jj})\sigma_{ji},
\end{eqnarray}
with $\rho(t)\equiv \rho$. Here, different energy levels are
assumed to have different dissipation channels. The operator
$\rho(t)$ is the reduced density matrix of the three-level SFQC.
We will study the steady-state response; thus, the thermal
equilibrium state $\overline{\rho}$ for $V(t)=0$ with matrix
elements $\overline{\rho}_{lj}$ is added to the master equation.
Also, $\gamma_{ii}$ is the pure dephasing rate of the energy level
$|i\rangle$, while $\gamma_{ij}=\gamma_{ji}$ (with $i\neq j$) are
the off-diagonal decay rates.

\section*{Sum- and difference-frequency generations}

We assume that the SFQC is in the thermal equilibrium state
$\overline{\rho}$ when $V(t)=0$. To study the steady-state
response of the three-level SFQC to weak external fields, we have
to obtain the solution of the reduced density matrix $\rho$ for
the three-level SFQC in Eq.~(\ref{eq:2}) by solving the following
equations:
\begin{eqnarray}
\dot{\rho}_{ij}(t)&=&\frac{1}{i\hbar}[H_{T},\rho(t)]_{ij}-\frac{1}{2}\Gamma_{ij}\widetilde{\rho}_{ij}(t), \,\, i\neq j,\nonumber \\
\dot{\rho}_{11}(t)&=&\frac{1}{i\hbar}[H_{T},\rho(t)]_{11}+\gamma_{12}\widetilde{\rho}_{22}(t)+\gamma_{13}\widetilde{\rho}_{33}(t),\nonumber \\
\dot{\rho}_{22}(t)&=&\frac{1}{i\hbar}[H_{T},\rho(t)]_{22}-\gamma_{12}\widetilde{\rho}_{22}(t)+\gamma_{23}\widetilde{\rho}_{33}(t),\nonumber\\
\dot{\rho}_{33}(t)&=&\frac{1}{i\hbar}[H_{T},\rho(t)]_{33}-(\gamma_{13}+\gamma_{23})\widetilde{\rho}_{33}(t)
\end{eqnarray}
with the parameters $\Gamma_{12}=\gamma_{12}$,
$\Gamma_{13}=\gamma_{13}+\gamma_{23}+\gamma_{33}$ and
$\Gamma_{23}=\gamma_{12}+\gamma_{13}+\gamma_{23}+\gamma_{22}+\gamma_{33}$,
derived from Eq.~(\ref{eq:2}). Note that
$\Gamma_{ij}=\Gamma_{ji}$. Here we define
$\widetilde{\rho}_{ij}(t)=\rho_{ij}(t)-\overline{\rho}_{ij}$.
Because the external fields are weak, the solution of $\rho(t)$
can be obtained by expressing $\rho(t)$ in the form of a
perturbation series in $V_{T}(t)$, i.e.,
\begin{equation}
\rho(t)=\rho_0+\rho_{1}(t)+\rho_{2}(t)+\cdots,
\end{equation}
with the density matrix operator $\rho_{0}=\overline{\rho}$ in the
zeroth-order approximation. We define the magnetic polarization
$P$ due to the external field as $P={\rm Tr}[\rho(t)I]$, in
analogy to the electric polarization~\cite{Shen-book}, then the
second-order magnetic polarization can be given as $P^{(2)}={\rm
Tr}[\rho_{2}(t)I]$, and then the second-order magnetic
susceptibility can be given by
\begin{equation}
\chi^{(2)}(\omega)=\frac{P^{(2)}(\omega)}{\Phi(\omega_{1})\Phi(\omega_{2})}.
\end{equation}
   In our study, since the
condition $|E_i-E_j|\gg k_{B}T$ (with $i\neq j$) is satisfied,
then the system is in its ground state $|1\rangle$ in the thermal
equilibrium state, i.e.,
$\rho_{0}=\overline{\rho}=|1\rangle\langle 1|$.

\subsection{Sum-frequency generation}

To study the microwave generation of the sum-frequency, we now
assume that the two external magnetic fluxes are applied to the
three-level SFQC. As schematically shown in Fig.~1(b), one
magnetic flux with frequency $\omega_{1}$ ($\omega_{2}$) induces
the transition between the energy levels $|1\rangle$ and
$|2\rangle$ ($|2\rangle$ and $|3\rangle$). In this case, the
interaction Hamiltonian $V_{T}(t)$ between the three-level SFQC
and the two external fields is given by
\begin{eqnarray}
V_{1}(t)=\sum_{i=1,2} I_{i,i+1}(f)\sigma_{i,i+1}\Phi(\omega_{i})
\exp(\I\omega_{i}t)+{\rm H.c.} \label{eq_V1}
\end{eqnarray}
under the rotating-wave approximation. On replacing $V_{T}(t)$ in
Eq.~(\ref{eq:2}) by $V_{1}(t)$, and using the perturbation theory
discussed above, we can obtain the reduced density matrix of the
three-level SFQC, up to second order in $V_{1}(t)$, and find the
second-order magnetic susceptibility as
\begin{eqnarray}\label{eq:7}
\chi^{(2)}(\omega_{+})=\frac{I_{12}(f)I_{23}(f)I_{31}(f)}
{(\I\omega_{1}-\I\omega_{21}+\Gamma_{21})(\I\omega_{+}-\I\omega_{31}+\Gamma_{31})}
\end{eqnarray}
for the sum-frequency generation with
$\omega_{+}=\omega_{1}+\omega_{2}$, and
$\omega_{ij}=(E_{i}-E_{j})/\hbar$, with $i>j$.
Equation~(\ref{eq:7}) obviously shows that the second-order
magnetic susceptibility is proportional to the product of the
three different electric dipole-like matrix elements (or
transition matrix elements) $I_{ij}(f)$, with $i \neq j$.
Therefore, for a given reduced magnetic flux $f$, the maximum
value of the susceptibility in Eq.~(\ref{eq:7}) is
$\chi^{(2)}_{\max}(\omega_{+})=I_{12}I_{23}I_{31}/(\Gamma_{21}\Gamma_{31})$,
when $\omega_{+}=\omega_{31}$ and $\omega_{1}=\omega_{21}$.

\subsection{Difference-frequency generation}

Similarly, the difference-frequency can also be generated by using
a three-level SFQC. We assume that a magnetic flux with frequency
$\omega_{1}$ ($\omega_{2}$) is applied between the energy levels
$|1\rangle$ and $|3\rangle$ ($|2\rangle$ and $|3\rangle$) as shown
in Fig.~1(c). In this case, the interaction between the
three-level SFQC and the external magnetic fields can be described
by
\begin{eqnarray}
V_{2}(t)=\sum_{i=1,2} I_{i,3}(f)\sigma_{i,3}\Phi(\omega_{i})
\exp(\I\omega_{i}t)+{\rm H.c.} \label{eq_V2}
\end{eqnarray}
under the rotating-wave approximation.

Using the same calculation as for Eq.~(\ref{eq:7}), we can also
obtain the second-order magnetic susceptibility  of the
difference-frequency $\omega_{-}=\omega_{2}-\omega_{1}$ as
\begin{eqnarray}\label{eq:8}
\chi^{(2)}(\omega_{-})=\frac{I_{13}(f) I_{21}(f)I_{32}(f)}
{(\I\omega_{-}-\I\omega_{21}+\Gamma_{21})(\I\omega_{1}-\I\omega_{31}+\Gamma_{31})}.
\end{eqnarray}
For a given reduced magnetic flux $f$, the maximum amplitude
$\chi^{(2)}_{\max}(\omega_{-})=I_{13}I_{21}I_{32}/(\Gamma_{21}\Gamma_{31})$
of the susceptibility in Eq.~(\ref{eq:8}) for the
difference-frequency can be obtained under the resonant driving
conditions: $\omega_{-}=\omega_{21}$ and $\omega_{1}=\omega_{31}$.

\begin{figure}
\begin{center}
\epsfig{file=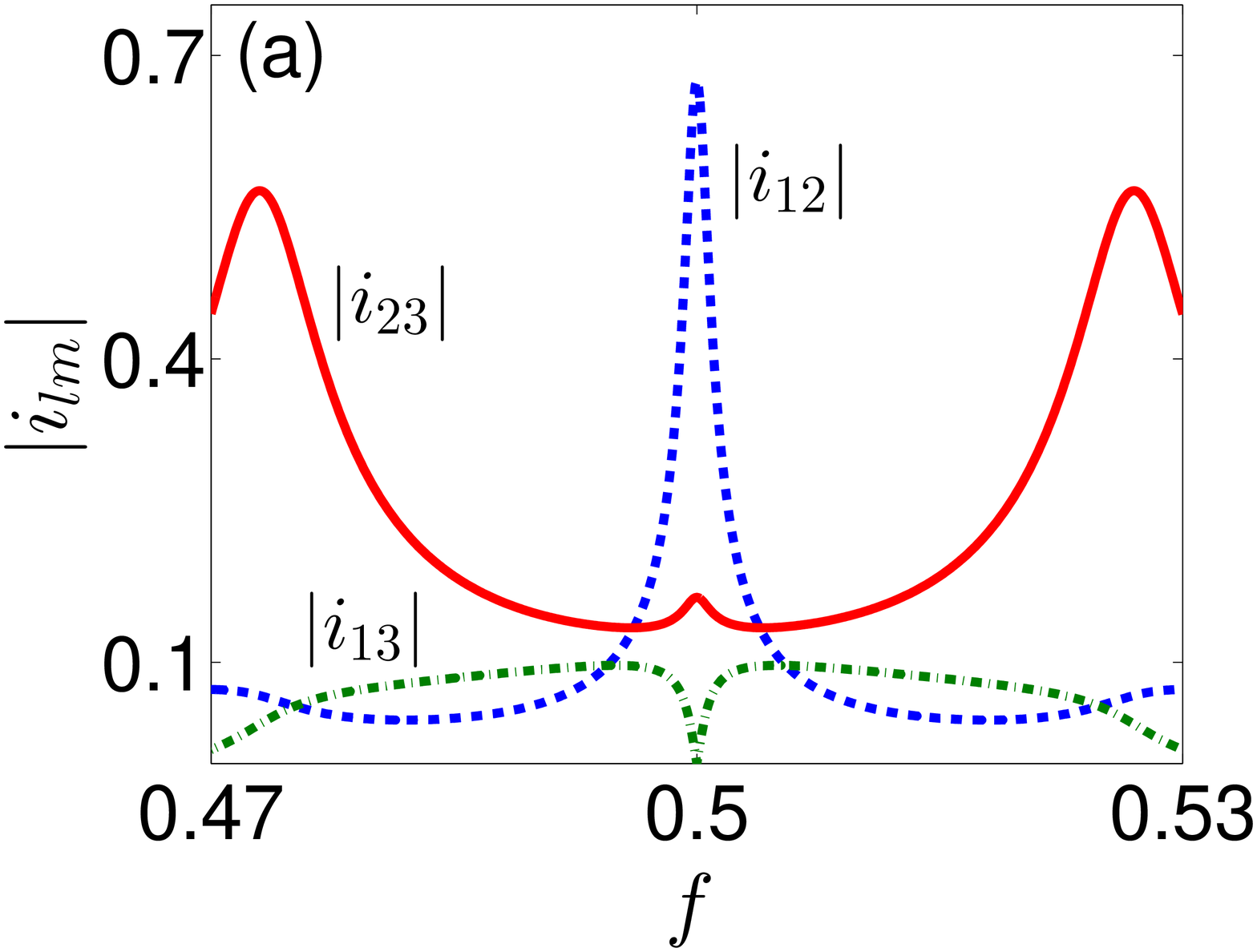,width=8cm}
\epsfig{file=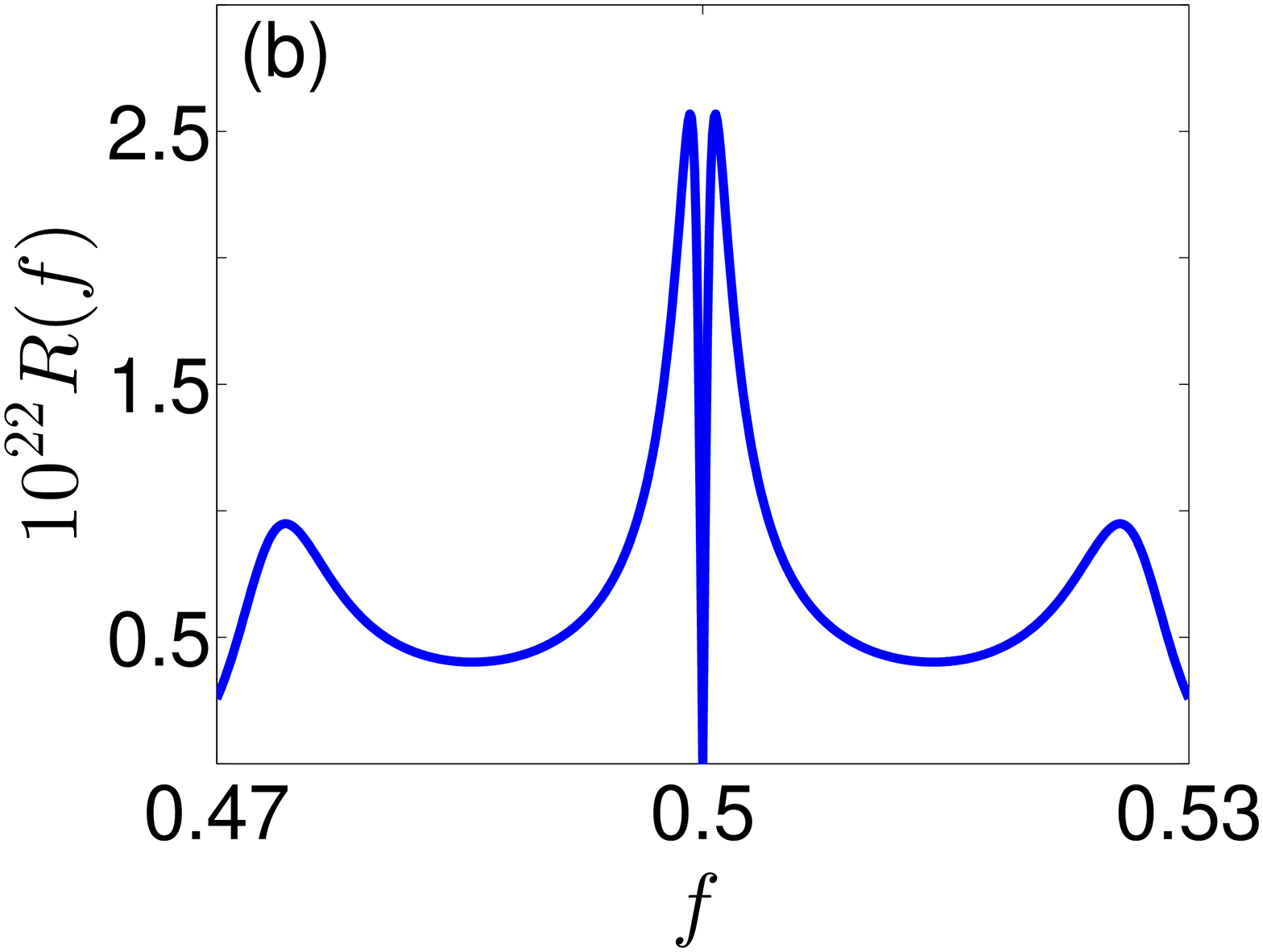,width=8cm}
\epsfig{file=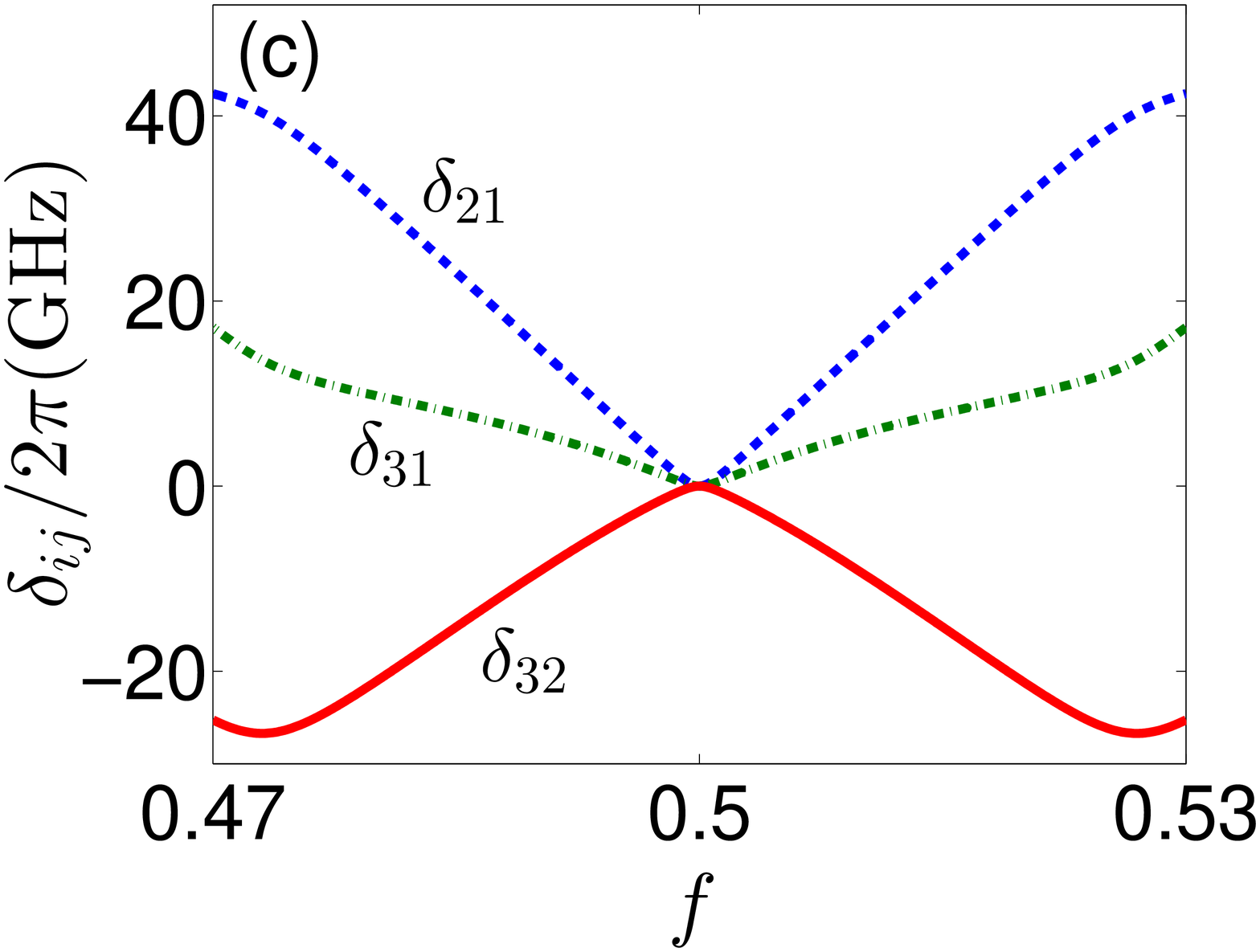,width=8cm}
\end{center}\caption{(a) Modulus of the renormalized transition
elements $i_{12}=I_{12}/I_{0}$,  $i_{23}=I_{23}/I_{0}$, and
$i_{13}=I_{13}/I_{0}$ versus the reduced magnetic flux $f$.
The modulus $R(f)$, given by Eq.~(\ref{eq_R}), and the
detuning $\delta_{ij}(f)=\omega_{ij}(f)-\omega_{ij}^{({\rm
opt})}$, versus $f$, are plotted in panels (b) and (c),
respectively. Here, $\omega_{ij}(f)$ ($\omega_{ij}^{({\rm
opt})}$) are the $f$-dependent transition frequencies
(transition frequencies at the optimal point with $f=0.5$)
between two different energy levels $|i\rangle$ and
$|j\rangle$ ($i>j$). The SFQC parameters are here taken as
$E_{J}/h=192$ GHz, $E_{J}/E_{c}=48$, and $\alpha=0.8$ with
$h$ being the Planck constant.} \label{fig2}
\end{figure}
\subsection{Numerical simulation}
Both Eqs.~(\ref{eq:7}) and~(\ref{eq:8}) show that the
susceptibilities of the sum- and difference-frequencies can be
controlled by the bias magnetic flux $\Phi_{e}$. According to the
analysis of the inversion symmetry for flux quantum
circuits~\cite{liu2005}, we know that the three-level SFQC has a
well-defined symmetry at the optimal point $f=0.5$ and it behaves
as natural three-level atoms with the $\Xi$-type (or ladder-type)
transition. In this case, the transition matrix elements between
the energy levels $|1\rangle$ and $|3\rangle$ is zero, i.e.,
$I_{13}(f=0.5)=I_{31}(f=0.5)=0$, and both susceptibilities,
$\chi^{(2)}(\omega_{+})$ in Eq.~(\ref{eq:7}) and
$\chi^{(2)}(\omega_{-})$ in Eq.~(\ref{eq:8}), are zero. Thus, the
microwave sum- or difference-frequencies cannot be generated at
the optimal point as for natural three-level atoms with the
electric-dipole selection rule. Equations~(\ref{eq:7}) and
(\ref{eq:8}) also tell us that the amplitudes of the
susceptibilities for both the sum- and difference-frequencies are
proportional to the modulus $R(f)$ of the product of the three
different transition matrix elements, i.e.,
\begin{equation}
  R(f)\equiv|I_{12}(f)I_{23}(f)I_{31}(f)|=|I_{21}(f)I_{32}(f)I_{13}(f)|.
\label{eq_R}
\end{equation}
Thus, the maximum value $R^{(\max)}(f)$ of $R(f)$ corresponds to
the maximal susceptibilities under the resonant driving condition.
To show clearly how the bias magnetic flux $\Phi_{e}$ can be used
to control the sum- and difference-frequency generations,   the
three  transition elements $|I_{12}|$, $|I_{23}|$ and $|I_{13}|$
versus the reduced magnetic flux $f$ are plotted in Fig.~2(a).
Also, the $f$-dependent product $|I_{12}I_{23}I_{31}|$ is plotted
in Fig.~2(b).  Here, we take experimentally accessible parameters,
for example, $\alpha=0.8$, $E_{J}/h=192$ GHz, and
$E_{J}/E_{c}=48$, where $E_{c}$ is the charging energy and $h$ is
the Planck constant. These data are taken from the RIKEN-NEC group
for their most recent, unpublished, experimental setup.
Figures~2(a) and 2(b) clearly show that the bias magnetic flux
$\Phi$. i.e., $f=\Phi/\Phi_{0}$, can be used to tune the
transition elements, and then $R(f)$ is also tunable. We find that
$R(f)$ is zero, at the optimal point corresponding to the zero
signal for the sum- and difference-frequency generations, because
the transition selection rule at this point makes the transition
element $I_{13}=0$, as shown in Fig.~2(a).  That is, the
transition between the energy levels $|1\rangle$ and $|3\rangle$
is forbidden. However, the sum- and difference-frequencies can be
generated when $f\neq 0.5$, and the maximum $R^{(\max)}(f)$
corresponds to two symmetric points with $ f=0.4992$ and
$f=0.5008$. To show the tunability of the frequency generation, we
now define a maximum variation $\delta^{({\rm max})}_{ij}$ (
$i>j$) of the sum- and difference-frequency generation as
\begin{equation}
\delta^{(\max)}_{ij}=\frac{1}{2\pi}(\omega_{ij}-\omega_{ij}^{({\rm
opt})})
\end{equation}
for a given range of the reduced magnetic flux $f$. Here,
$\omega_{ij}^{({\rm opt})}$ denotes the transition frequency
between the energy levels $|i\rangle$ and $|j\rangle$ at the
optimal point.

Figure~2(c) shows that the maximum variation
$\delta^{(\max)}_{31}$ of the sum-frequency is
$\delta^{(\max)}_{31}=(\omega_{31}-\omega_{31}^{({\rm
opt})})/(2\pi)\approx 17$ GHz for $0.5<f<0.53$. However, the
maximum variation $\delta^{(\max)}_{21}$ or $\delta^{(\max)}_{32}$
of the difference-frequency is
$\delta^{(\max)}_{21}=(\omega_{21}-\omega_{21}^{({\rm
opt})})/(2\pi)\approx 42$ GHz or
$\delta^{(\max)}_{32}=(\omega_{32}-\omega_{32}^{({\rm
opt})})/(2\pi)\approx 26$ GHz for $0.5<f<0.53$. Thus, the
tunability for the sum- and difference-frequency generations can
be, in principle, over a very wide GHz range, by using the bias
magnetic flux $\Phi_{e}$.

\begin{figure}
\begin{center}
\epsfig{file=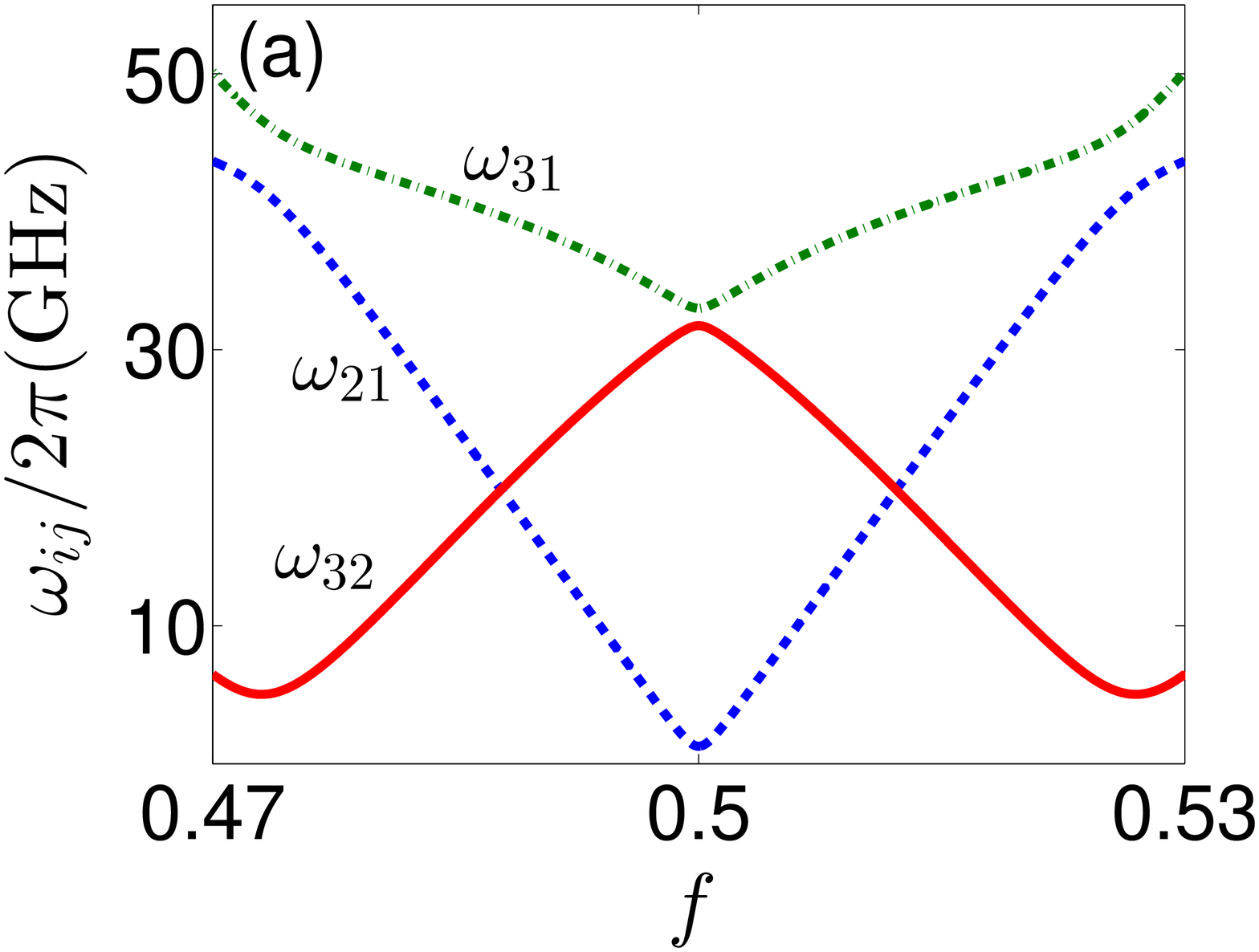,width=8cm}
\epsfig{file=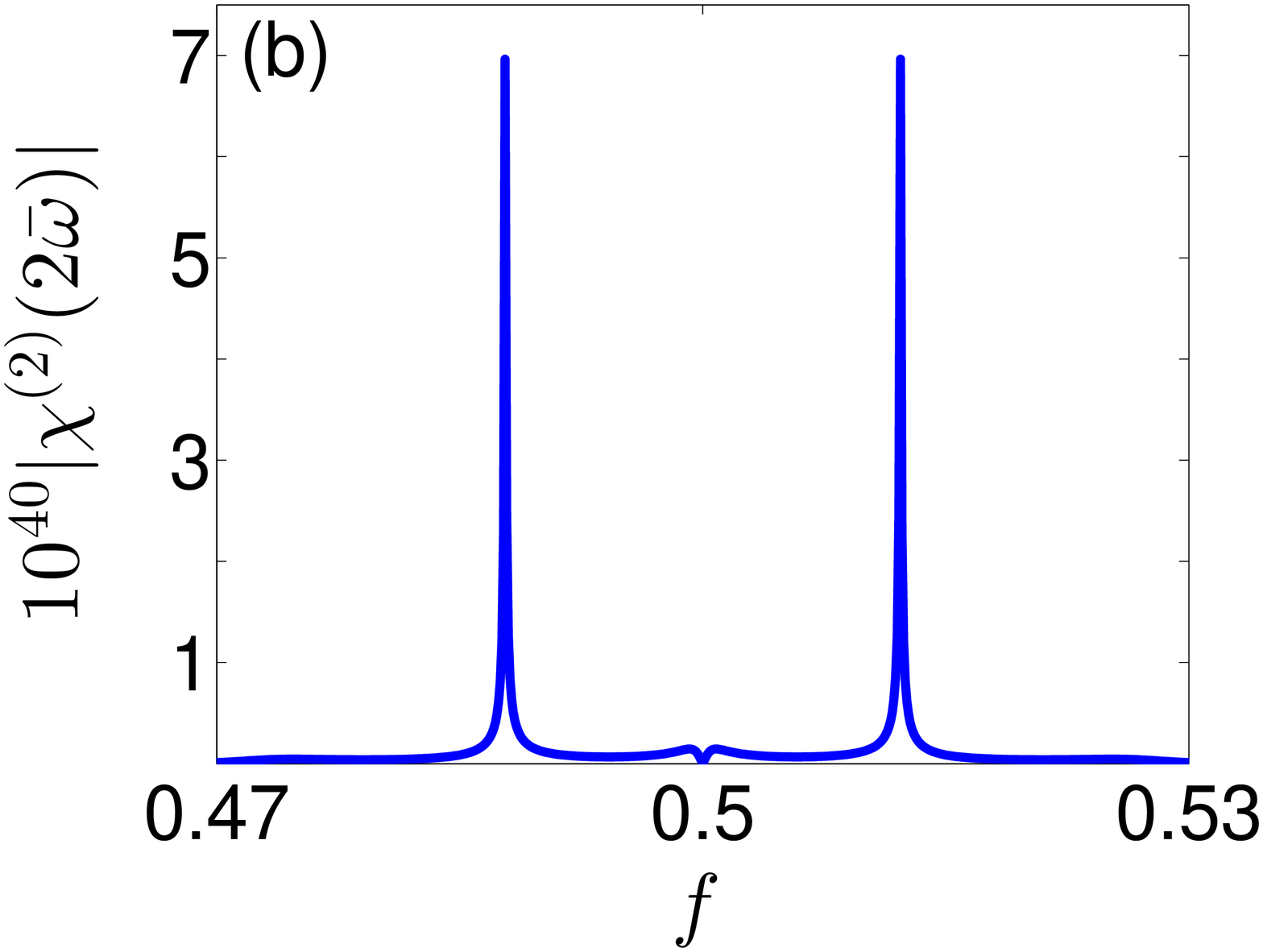,width=8cm}
\end{center}\caption{(a) Three transition frequencies $\omega_{ij}$
($i>j$) versus  the reduced magnetic flux $f$ is plotted. The
crossing points for the curves of $\omega_{21}$ and
$\omega_{32}$ correspond to $\omega_{21} =\omega_{32}$. (b)
The amplitude $|\chi^{(2)}(2\overline{\omega})|$ of the
susceptibility in Eq.~(\ref{eq:20}) versus $f$ is plotted
with, e.g., $\Gamma_{21}/2\pi=50$ MHz and
$\Gamma_{31}/2\pi=30$ MHz. The same SFQC parameters as in
Fig.~2 are used in both panels.} \label{fig3}
\end{figure}
\section*{Second-harmonic generation}

From Eqs.~(\ref{eq:7}) and~(\ref{eq:8}), we find that the
second-harmonic and zero-frequency signals can also be generated
in three-level SFQCs when two applied external fields have the
same frequency and satisfy the condition
\begin{eqnarray}
  \omega_{1}=\omega_{2}= {\textstyle\frac12} \omega_{31}=\overline{\omega}.
\label{eq_SHG}
\end{eqnarray}
Let us now discuss  second-harmonic generation. As shown in
Fig.~3(a), we can find two values of the reduced magnetic flux,
$f=0.4878$ or $f=0.5122$, such that
$\omega_{31}=2\omega_{21}=2\omega_{32}$.  In this case, the
susceptibility of the second harmonic reaches its maximum, when an
external field with the same frequency as
$\omega_{21}=\omega_{32}$ is applied to the three-level SFQC.
However, the second-order susceptibility becomes small when the
magnetic field deviates from the points $f=0.4878$ or $f=0.5122$
because of the  anharmonicity of the energy-level structure for
the SFQC. If we assume that the anharmonicity is characterized by
\begin{equation}
\delta(f)=\overline{\omega}(f)-\omega_{21}(f)=\frac{\omega_{31}(f)}{2}-\omega_{21}(f),
\end{equation}
then the second-order susceptibility for the second-harmonic
generation can be approximately written as
\begin{equation}\label{eq:9}
\chi^{(2)}(2\overline{\omega})=
\frac{I_{12}(f)I_{23}(f)I_{31}(f)}{\left[i\delta(f)+\Gamma_{12}\right]\Gamma_{13}}.
\end{equation}
We note that this equation for the second-order susceptibility
$\chi^{(2)}(2\overline{\omega})$ is a rough approximation when
$\overline{\omega}(f)=\omega_{21}(f)$, i.e., $\delta=0$. Because
the independent-environment assumption for the decays of different
energy levels might not always hold and the dissipation rates
$\Gamma_{12}$ and $\Gamma_{13}$ should be modified. However, the
main physics is not changed. In Fig.~3(b), as an example, the
amplitude of $\chi^{(2)}(2\overline{\omega})$, which is given by
\begin{equation}\label{eq:20}
|\chi^{(2)}(2\overline{\omega})|=\frac{|I_{12}(f)I_{23}(f)I_{31}(f)|}{\Gamma_{13}
\sqrt{\delta^2+\Gamma^2_{12}}},
\end{equation}
is plotted as a function of $f$ for given parameters, e.g.,
$\Gamma_{12}/2\pi=50$ MHz  and $\Gamma_{13}/2\pi=30$ MHz. It
clearly shows that the maximum amplitude of the susceptibility
$\chi^{(2)}(2\overline{\omega})$ corresponds to the reduced
magnetic flux $f=0.4878$ or $f=0.5122$, in which the three energy
levels have a harmonic structure. It should be noted that we take
$\Gamma_{21}$ and $\Gamma_{31}$ as the $f$-independent parameters
for convenience when Fig.~3(b) is plotted. In practice, they
should also depend on $f$.

\begin{figure}
\begin{center}
\epsfig{file=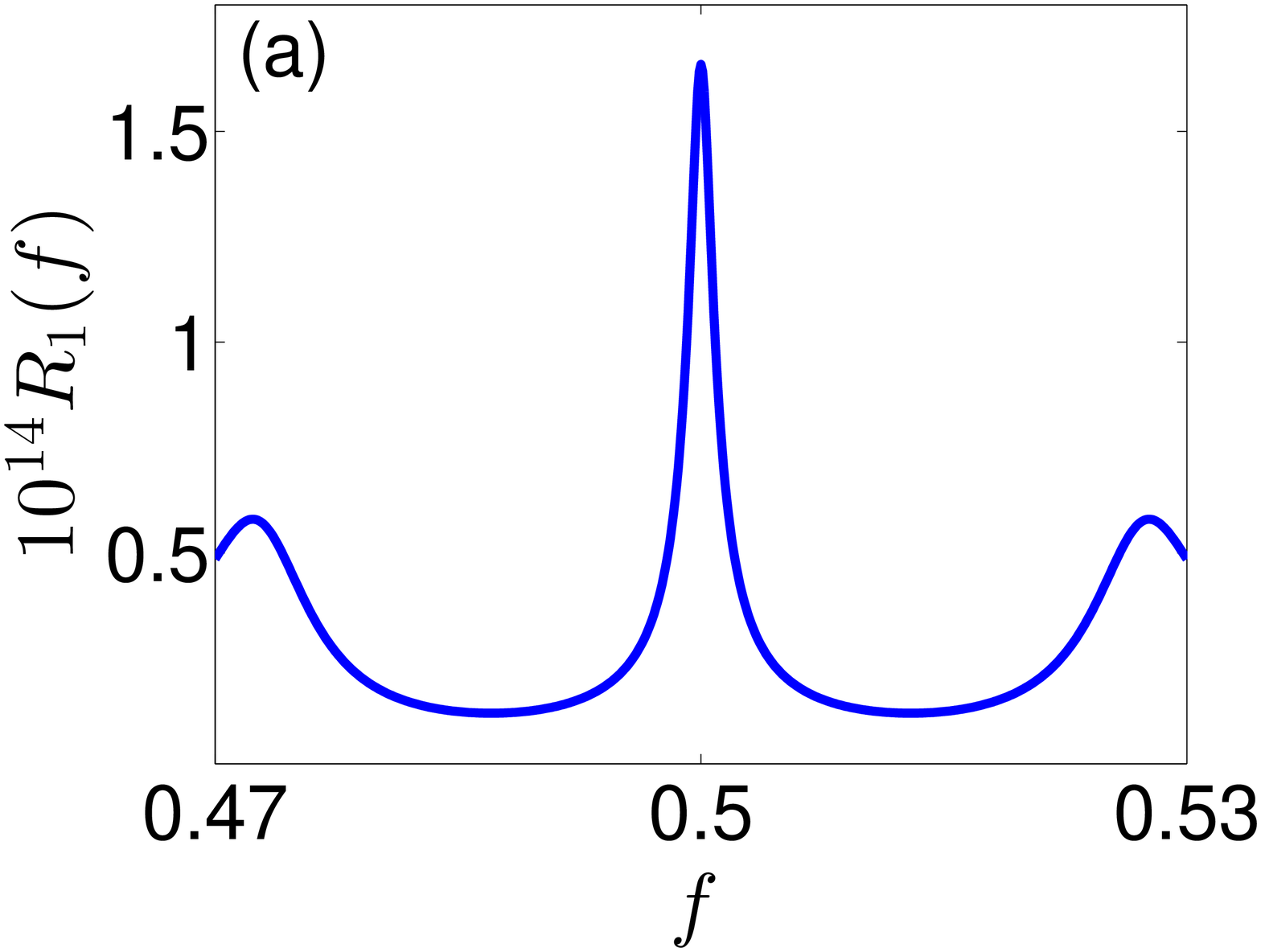,width=8cm}
\epsfig{file=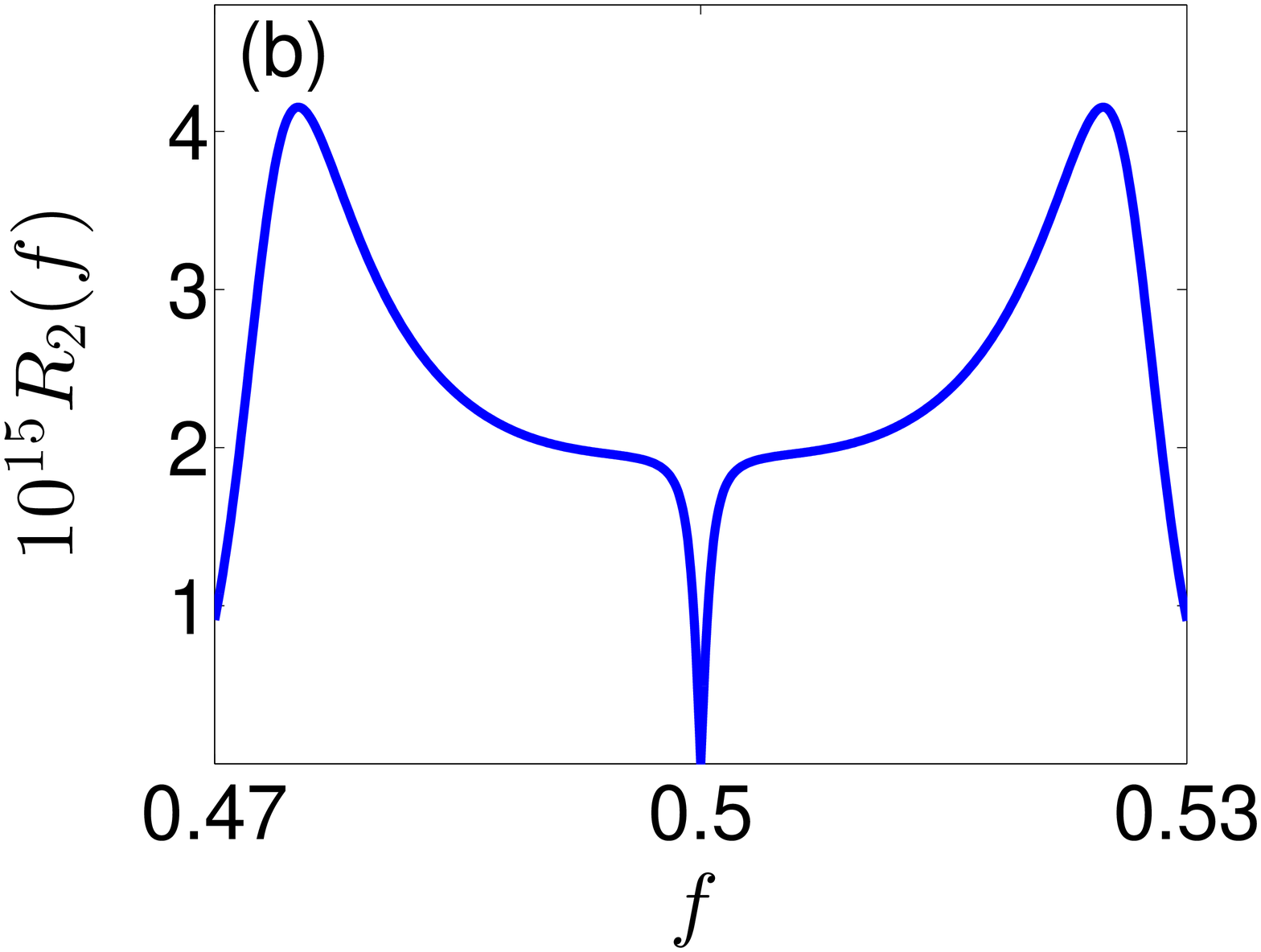,width=8cm}
\end{center}\caption{Moduli
$R_{1}(f)\equiv|I_{21}(f)I_{32}(f)|$ and
$R_{2}(f)\equiv |I_{13}(f)I_{32}(f)|$ versus the reduced magnetic flux $f$ are plotted in (a) and (b), respectively, for the same SFQC parameters as
in Fig.~2.} \label{fig4}
\end{figure}
\section*{Measurements}

We now take the sum-frequency generation as an example to show how
to measure the frequency generation by coupling the three-level
SFQC to the continuum of electromagnetic modes confined in a 1D
transmission line as for measuring the resonance fluorescence of
single artificial atoms~\cite{Astatiev,LanZhou}. As discussed in
Ref.~\cite{Blais2010}, if the three transition frequencies of the
three-level SFQC are much larger than the decay rates, then we can
consider that the decays of different energy levels occur via
different dissipation channels. In this case, the interaction
Hamiltonian between the three-level SFQC and the continuum modes
in the transmission line can be modeled as
\begin{eqnarray}
H_{\rm in}&=&\int_{-\infty}^{\infty}\frac{d\omega}{\sqrt{2\pi}}
\left[\sqrt{\gamma_{12}}a^{\dagger}(\omega)\sigma_{12}
+\sqrt{\gamma_{23}}b^{\dagger}(\omega)\sigma_{23}\right.\nonumber\\
&&+\left.\sqrt{\gamma_{13}}c^{\dagger}(\omega)\sigma_{13}\right]+{\rm
H.c.}
\end{eqnarray}
under the Markovian approximation with the bosonic commutation
relation $[\alpha(\omega), \beta^{\dagger}(\omega^{\prime})]
=\delta_{\alpha,\beta}\delta(\omega-\omega^{\prime})$  with
$\alpha,\;\beta=a,\,b,\,c$ for the three kinds of different
continuum mode operators. According to the input-output
theory~\cite{Walls-book}, the output field centered at the
sum-frequency $\omega_{1}+\omega_{2}=\omega_{+}$ can be given as
\begin{eqnarray}
  \langle c_{{\rm out}}(t)\rangle=\langle c_{\rm
in}(t)\rangle+\sqrt{\gamma_{13}}\langle\sigma_{13}(t)\rangle,
\label{eq_c}
\end{eqnarray}
since ${\rm Tr}[\rho\sigma_{13}(t)]={\rm
Tr}[\rho(t)\sigma_{13}]=\rho_{31}(t)$. Therefore, up to second
order in $V_{1}(t)$ for the sum-frequency generation, we can
approximately obtain the output of the sum-frequency generation as
\begin{equation}\label{eq:10}
\langle c_{{\rm out}}(t)\rangle =
-\frac{\sqrt{\gamma_{13}}}{\hbar^2}\frac{I_{21}(f)I_{32}(f)\Phi(\omega_{1})\Phi(\omega_{2})
\exp(-i\omega_{+} t)} {(i\omega_{21}-i\omega_{1}+\Gamma_{21})
(i\omega_{31}-i\omega_{+}+\Gamma_{31})}, 
\end{equation}
where the input field for the continuum mode $c(\omega)$ is in the
vacuum. Equation~(\ref{eq:10}) shows that the amplitude of the
output field is proportional to the intensities
$|\Phi(\omega_{1})|$ and $|\Phi(\omega_{2})|$ of the two external
magnetic fields, the modulus of the product of two transition
matrix elements $I_{21}(f)$ and $I_{32}(f)$, and the square root
of the decay rate $\gamma_{13}$. It is obvious that the intensity
of the output field can be tuned by the bias magnetic flux
$\Phi_{e}$. Similarly, the amplitude of the output field for the
difference-frequency generation described in Eq.~(\ref{eq:8}) is
proportional to the modulus of the product of two transition
matrix elements $I_{13}(f)$ and $I_{32}(f)$. The moduli
$R_{1}(f)\equiv|I_{21}(f)I_{32}(f)|$ and $R_{2}(f)\equiv
|I_{13}(f)I_{32}(f)|$ versus $f$ are plotted in Figs.~4(a) and
(b), which show that the amplitude of the output fields for the
sum- and difference-frequency generations can also be tuned by
$f$. However, the maximum value, corresponding to maximum
second-order susceptibility under resonant condition,  of
$R^{(\max)}(f)$ does not correspond to the maximum value of
$R_{1}(f)$ for the sum-frequency, or $R_{2}(f)$ for the
difference-frequency.

\section*{Conclusions}

We have proposed and studied a controllable method for generating
sum- and difference- frequencies by using three-wave mixing in a
single three-level SFQC driven by two weak external fields. Thus,
in perturbation theory, the noise and frequency shifts introduced
by the driving fields can be neglected and we can obtain all the
response functions of different frequencies. We point out that the
three-wave-mixing signal can only be generated when the inversion
symmetry of the potential energy for the SFQC is broken, that is,
the SFQC cannot work at the optimal point. Otherwise, the
transition between the ground state and the second-excited state
is forbidden, so three-wave mixing cannot be generated as in
natural-atom systems. We have shown that the generated microwave
signal can be tuned in a very large GHz range. We have also
discussed how to generate second-harmonics in the single SFQC. We
note that three-wave mixing can also occur in superconducting
phase~\cite{Martinis,Hakonen1,Hakonen2} and
transmon~\cite{transmon} qutrits, when the inversion symmetry of
their potential energies is broken. In particular, the phase
qutrits might be better for second-harmonic generation because of
their small anharmonicity. It should be pointed out that the
microwave signal with the sum-frequency might exceed the
high-frequency cutoff of the cryogenic amplifier~\cite{Astatiev}.
Thus, the difference-frequency generation should be easier to be
experimentally accessed.

In contrast to Ref.~\cite{Roch-PRL},  with a frequency tunability
of about 500 MHz, we show that the tunability of the output
frequency using single flux qubit circuits can be a few GHz. Our
proposal is valid not only for nondegenerate three-wave mixing,
but it can also be applied for second-harmonic generation by
changing the bias magnetic flux. Also, contrary to
Ref.~\cite{Roch-PRL} , where the circuit itself is in the
classical regime,  in our study, the three-wave mixing is
generated using excitations of real quantized energy levels of the
artificial atoms. Such excitation will result in a strong
nonlinearity. Thus, the three-wave mixing in single artificial
atoms can be used to generate entangled microwave photons and act
as entanglement amplifier or correlated lasing. These could be
important toward  future quantum networks.

In summary,  our study could help generating three- or multi-wave
mixing using single artificial atoms.  The proposed method is
simple  and could be used for  manipulating second-order and other
nonlinear processes in the microwave regime by using single
superconducting artificial atoms.  Our proposal is realizable
using current experimental parameters of superconducting flux
qubit circuits.



\noindent {\bf Acknowledgement:} Y.X.L. is supported by the
National Basic Research Program of China Grant
No.~2014CB921401, the NSFC Grants No.~61025022, and
No.~91321208. A.M. is supported by Grant No.
DEC-2011/03/B/ST2/01903 of the Polish National Science
Centre. F.N. is partially supported by the RIKEN iTHES
Project, MURI Center for Dynamic Magneto-Optics, and a
Grant-in-Aid for Scientific Research (S). Z.H.P. and J.S.T.
were supported by Funding Program for World-Leading
Innovative R \& D on Science and Technology (FIRST), MEXT
KAKENHI ``Quantum Cybernetics''.

\noindent {\bf Correspondence and requests for materials}
should be addressed to Y.X.L. (yuxiliu@mail.tsinghua.edu.cn)

\noindent {\bf Author contributions}: Y.X.L. proposed the
main idea. Y.X.L., H.C.S., Z.H.P., A.M. and F.N. contributed
to the findings of this work and wrote the manuscript. J.S.T.
participated in the discussions.

\noindent {\bf Additional information:} The authors declare
that they have no competing financial interests.

\end{document}